\documentclass[aps,pre,twocolumn,superscriptaddress,amsmath,amssymb,floatfix,
showpacs]{revtex4}
\usepackage{graphicx}
\usepackage{times}
\begin{document}
\def\d{\delta}
\def\D{\Delta}
\def\s{\sigma}
\def\g{\gamma}
\def\e{\epsilon}
\def\b{\beta}
\def\a{\alpha}

\title{Duplication-divergence model of protein interaction network
}
\author{I. Ispolatov}
\email{iispolat@lauca.usach.cl}
\altaffiliation{Permanent address:
Departamento de Fisica, Universidad de Santiago de Chile,
Casilla 302, Correo 2, Santiago, Chile}
\affiliation{Ariadne Genomics Inc,. Rockville, MD 20850}
\author{P. L. Krapivsky}
\email{paulk@bu.edu}
\affiliation{Center for Polymer Studies and
Department of Physics, Boston University, Boston, MA 02215}
\author{A. Yuryev}
\email{ayuryev@ariadnegenomics.com}
\affiliation{Ariadne Genomics Inc,. Rockville, MD 20850}

\date{\today}
\begin{abstract}
  We show that the protein-protein interaction networks can be surprisingly
  well described by a very simple evolution model of duplication and
  divergence. The model exhibits a remarkably rich behavior depending on a
  single parameter, the probability to retain a duplicated link during
  divergence.  When this parameter is large, the network growth is not
  self-averaging and an average vertex degree increases algebraically. The
  lack of self-averaging results in a great diversity of networks grown out
  of the same initial condition.  For small values of the link retention
  probability, the growth is self-averaging, the average degree increases
  very slowly or tends to a constant, and a degree distribution has a
  power-law tail.

\end {abstract}

\maketitle

\section{Introduction}
A single- and multi- gene duplication plays crucial role in evolution
\cite{japan,bio}.  On the proteinomic level, the gene duplication leads to a
creation of new proteins that are initially identical to the original ones.
In a course of subsequent evolution, the majority of these new proteins are
lost as redundant, while some of them survive by diverging, i.e.  quickly
loosing old and possibly slowly acquiring new functions.

\begin{figure}
\includegraphics[width=.35\textwidth]{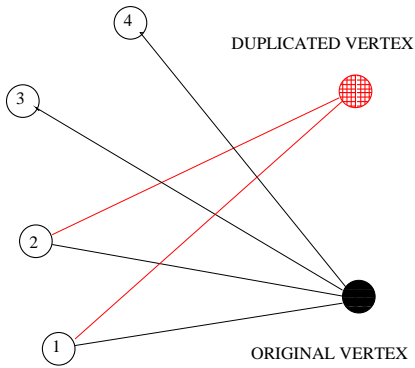}
\caption{\label{fig_uno}
A sketch of duplication and divergence event. Links between the 
duplicated vertex and vertices 3 and 4 disappeared as a result of divergence.} 
\end{figure}

The protein-protein interaction network is commonly defined as an evolving
graph with nodes and links corresponding to proteins and their interactions.
Thus a successful single-gene duplication event results in a creation of a
new node which is initially linked to all the neighbors of the original node.
Later, some links between each of the duplicates and their neighbors
disappear, Fig.~(\ref{fig_uno}).  Such network evolution process is commonly
called a duplication and divergence \cite{bio,wag}.  Although duplication and
divergence is usually considered as the growth mechanism only for
protein-protein networks, it also may play a role in a creation of certain
new nodes and links in the world wide web, growth of various networks of
human contacts by introduction of close acquaintances of existing members,
and evolution of many other non-biological networks.

Does the evolution dominated by duplication and divergence define the
structure and other properties of a network? So far, most of the attention
has been attracted to the study of a degree distribution $n_k$, which is a
probability for a vertex to have $k$ links.  Wagner \cite{wag} has provided a
numerical evidence that duplication-divergence evolution does not noticeably
alter the initial power-law degree distribution, provided that the evolution
is initiated with a fairly large network.  A somewhat idealized case of the
completely asymmetric divergence \cite{wag,wagas} when links are removed only
from one of the duplicates (as in Fig.~\ref{fig_uno}) was investigated in
Refs.~\cite{korea,chung}.  It was found that the emerging degree distribution
has a power-law tail: $n_k\sim k^{-\g}$ for $k\gg 1$.  Yet apart from the
shape of the degree distribution, a number of other perhaps even more
fundamental properties of duplication-divergence networks remain unclear:
\begin{enumerate}
\item How well does the model describe its natural prototype, the  
protein-protein networks\,? 
\item Is the total number of links a self-averaging quantity\,? 
\item How does the average total number of links depend on the network size
  $N$\,? 
\item Does the degree distribution scale linearly with $N$\,?
\end{enumerate}
A non-trivial answer to any of these questions would be much more
important than details of the tail of the degree distribution; the reason why
only these details are usually studied is that the more fundamental questions
are assumed to have trivial answers.

Here we shall attempt to answer above questions and we shall also look again
at the degree distribution of the duplication-divergence networks. As in
\cite{korea}, we consider a simple scenario of totally asymmetric divergence,
where evolution is characterized by a single parameter, link retention
probability $\s$. It turns out that even such idealized model describes the
degree distribution found in the biological protein-protein networks very
well.  We find that, depending on $\s$, the behavior of the system is
extremely diverse: When more than a half of links are (on average) preserved,
the network growth is non-self-averaging, the average degree diverges with
the network size, and while a degree distribution has a scaling form, it does
not resemble any power law.  In a complimentary case of small $\s$ the growth
is self-averaging, the average degree tends to a constant, and a degree
distribution approaches a scaling power-law form.

In the next section we formally define the model and compare the simulated
degree distribution to the observed ones.  The properties of the model are
first analyzed in the tractable $\s=1$ and $\s\rightarrow +0$ limits
(Sec.~III) and then in the general case $0<\s<1$ (Sec.~IV).  Section V gives
conclusions.

\section{duplication and divergence}

To keep the matter as simple as possible, we focus on the completely
asymmetric version of the model of duplication and divergence network
growth. The model is defined as follows (Fig.\ref{fig_uno}):
\begin{enumerate}
\item {\bf Duplication}. A randomly chosen target node is duplicated, that is
  its replica is introduced and connected to each neighbor of the target
  node.
\item {\bf Divergence}. Each link emanating from the replica is activated
  with probability $\sigma$ (this mimics link disappearance during
  divergence). If at least one link is established, the replica is
  preserved; otherwise the attempt is considered as a failure and the network
  does not change.  (The probability of the failure is $(1-\sigma)^k$ if the
  degree of the target node is equal to $k$.)
\end{enumerate}
In contrast to duplication-mutation models (see e.g.
\cite{korea,pastor,bb,cb,kd}), no new links are introduced.  Initial
conditions apparently do not affect the structure of the network when it
becomes sufficiently large; in the following, we always assume that the
initial network consists of two connected nodes. As in the observed
protein-protein interaction networks, in this model each node has at least
one link and the network remains connected throughout the evolution. These
features is the main distinction between our model and earlier models (see
e.g.  \cite{korea}) which allowed an addition of nodes with no links and
generated disconnected networks with questionable biological relevance.

\begin{figure}
\includegraphics[width=.45\textwidth]{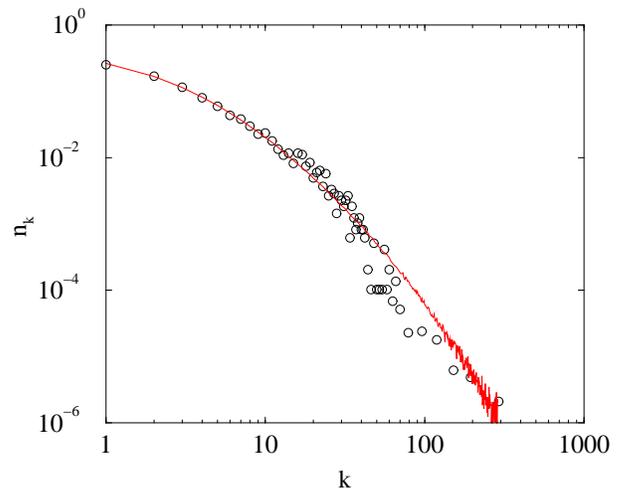}
\caption{\label{fig_yeast}
Degree distribution of protein-protein binding network of yeast with
$N_p$=4873 proteins and average degree $\langle d \rangle \approx 6.6$.
The link retainment probability of fitted simulated network $\s\approx0.413$.}
\end{figure}

\begin{figure}
\includegraphics[width=.45\textwidth]{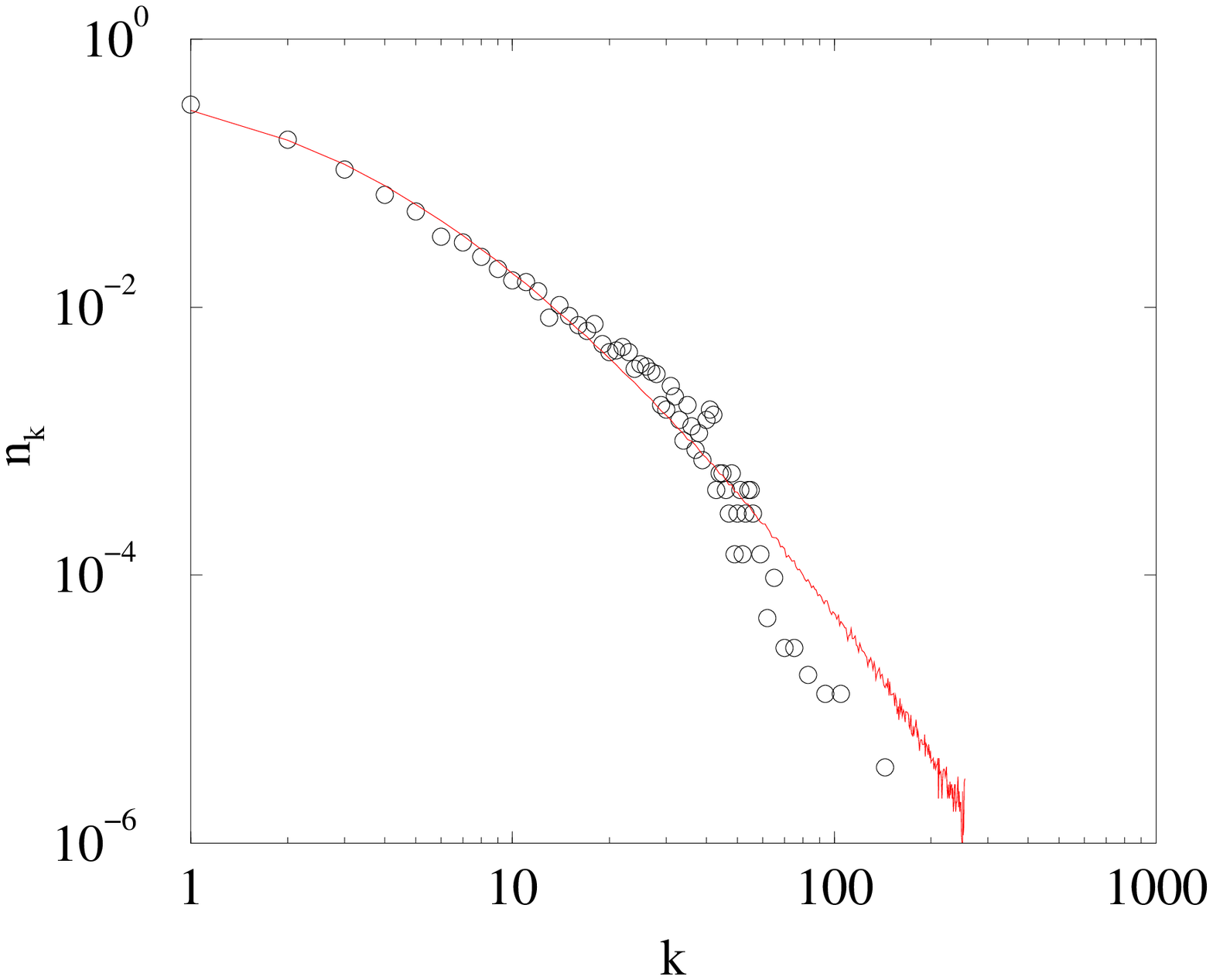}
\caption{\label{fig_fly}
Degree distribution of protein-protein binding network of fly with
$N_p$=6954 proteins and average degree $\langle d \rangle \approx 5.9$.
The link retainment probability of fitted simulated network $\s\approx0.380$.}
\end{figure}

\begin{figure}
\includegraphics[width=.45\textwidth]{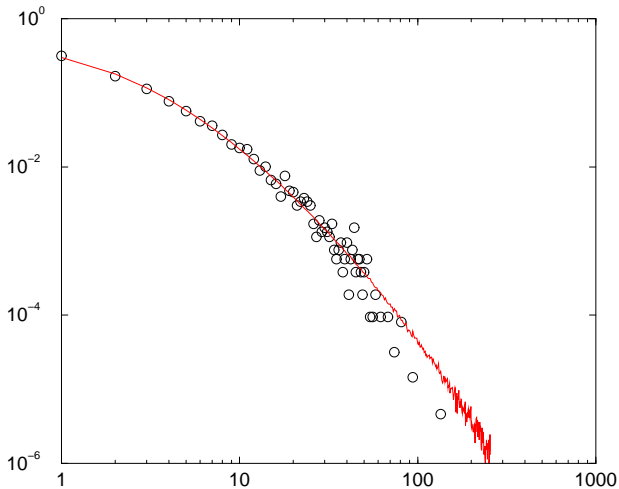}
\caption{\label{fig_human}
Degree distribution of protein-protein binding network of human with
$N_p$=5275 proteins and average degree $\langle d \rangle \approx 5.7$.
The link retainment probability of fitted simulated network $\s\approx0.375$.}
\end{figure}

The above simple rules generate networks which are strikingly similar to the
naturally occurring ones.  This is evident from
Figs.~\ref{fig_yeast}--\ref{fig_human} which compare the degree distribution of
the simulated networks and protein-protein binding networks of baker yeast,
fruit fly, and human.  The protein interaction data for all three species
were obtained from the Biological Association Network databases available
from Ariadne Genomics \cite{ag}.  The data for human ({\it H. sapiens})
protein network was derived from the Ariadne Genomics ResNet database
constructed from the various literature sources using Medscan
\cite{medscan1}.  The data for baker yeast ({\it S.  cerevisiae}) and fruit
fly ({\it D. melanogaster}) networks were constructed by combining the data
from published high-throughput experiments with the literature data obtained
using Medscan as well \cite{pwa}.

Each simulated degree distribution was obtained by averaging over 500
realizations. The values of the link retention probability $\s$ of simulated
networks were selected to make the mean degree $\langle d \rangle $ of the
simulated and observed networks equal. The number of nodes and the number of
links in the corresponding grown and observed networks were therefore equal
as well.

\begin{figure}
\includegraphics[width=.45\textwidth]{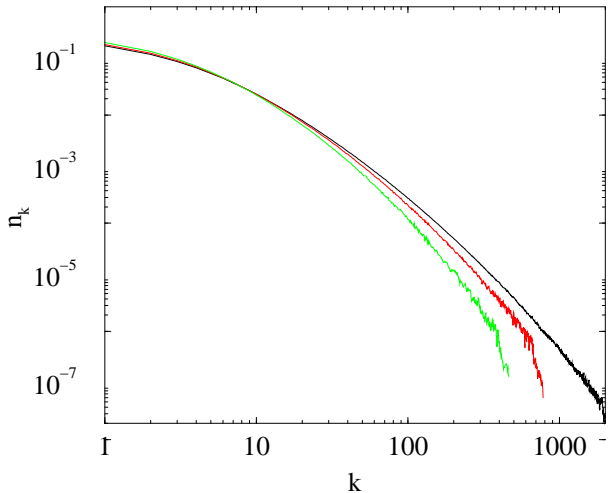}
\caption{\label{fig_simul}
Degree distributions of grown networks with (bottom to top) 
$10^4$, $10^5$, and $10^6$ vertices. The link retention probability $\s=0.45$,
all data was averaged over 100 realizations.} 
\end{figure}

Figures \ref{fig_yeast}--\ref{fig_human} demonstrate that even the most
primitive form of the duplication and divergence model (which does not
account for disappearance of links from the original node, introduction of
new links, removal of nodes, and many other biologically relevant processes)
reproduces the observed degree distributions rather well.  These figures also
show that the degree distributions of both simulated and naturally
occurring networks are not exactly resembling power-laws that they are
commonly fitted to (see, for example, \cite{wag}). A possible explanation is
that the protein-protein networks (naturally limited to few tens thousand of
nodes) are not large enough for a degree distribution to converge to its
power-law asymptotics. To probe the validity of this argument we present
(Fig.~\ref{fig_simul}) the degree distributions for networks of up to $10^6$
vertices with link retention probability similar to the fitted to the
observed networks, $\s=0.45$. It follows that a degree distribution does not
attain a power-law form even for very large networks, at least for naturally
occurring $\s\lesssim 1/2$.

\section{Solvable limits}

Here we analyze duplication-divergence networks in the limits $\sigma=1$ and
$\sigma\to 0$ when the model is solvable and (almost) everything can be
computed analytically.

\subsection{No divergence ($\sigma=1$)}

This case has already been investigated in Refs.~\cite{korea,chung,raval}.
Here we outline its properties as it will help us to pose relevant questions
in the general case when divergence is present.

When $\sigma=1$, each duplication attempt is successful and the network
remains a complete bipartite graph throughout the evolution: Initially it is
$K_{1,1}$; at the next stage the network turns into $K_{2,1}$ or $K_{1,2}$,
equiprobably; and generally when the number of nodes reaches $N$, the network
is a complete graph $K_{j,N-j}$ with every value $j=1,\ldots,N-1$ occurring
equiprobably. In the complete bipartite graph $K_{j,N-j}$ the degree of a
node has one of the two possible values: $j$ and $N-j$. Hence in any
realization of a $\s=1$ network, the degree distribution is the sum of two
delta functions: $N_k(j)=j\delta_{k,N-j}+(N-j)\delta_{k,j}$.  Averaging over
all realizations we obtain
\begin{equation}
\label{Nkav}
\langle N_k\rangle=\frac{1}{N-1}\sum_{j=1}^{N-1}N_k(j)=
\frac{2(N-k)}{N-1}
\end{equation}
The total number of links $L$ in the complete graph $K_{j,N-j}$  is $L=j(N-j)$.
Averaging over all $j$ we can compute any moment $\langle L^p\rangle$; for
instance, the mean is equal to
\begin{equation}
\label{Lav}
\langle L\rangle=\frac{1}{N-1}\sum_{j=1}^{N-1}j(N-j)=\frac{N(N+1)}{6}
\end{equation}
and the mean square is given by
\begin{equation}
\label{Lav2}
\langle L^2\rangle=\frac{N(N+1)(N^2+1)}{30}
\end{equation}

In the thermodynamic limit $N\to\infty, L\to\infty$, the link distribution
$P_N(L)$ becomes a function of the single scaling variable $\ell=L/N^2$,
namely:
\begin{equation}
\label{PNL}
P_N(L)=\frac{1}{N-1}\sum_{j=1}^{N-1}\d_{L,j(N-j)}
\to N^{-2}{\cal P}(\ell)
\end{equation}
with ${\cal P}(\ell)=2/\sqrt{1-4\ell}$. The key feature of the networks
generated without divergence ($\sigma=1$) is the lack of self-averaging. In
other words, fluctuations do not vanish in the thermodynamic limit. This is
evident from Eqs.~(\ref{Lav})--(\ref{PNL}): In the self-averaging case we
would have had $\langle L^2\rangle/\langle L\rangle^2=1$ (instead of the
actual value $\langle L^2\rangle/\langle L\rangle^2=6/5$) and the scaling
function ${\cal P}(\ell)$ would be the delta function. 
The lack of self-averaging implies that the future is uncertain --- a few
first steps of the evolution drastically affect the outcome.

Finally we mention that the $\sigma=1$ limit of our model
is equivalent to the classical P\'olya's urn model \cite{pol}.
The urn models have been studied in the probability theory \cite{JK}, have
applications ranging from biology \cite{life} to computer
science \cite{BP,A}, and remain in the focus of the current research (see
e.g. \cite{KMR,Fla} and references therein).

\subsection{Maximal divergence ($\sigma=+0$)}

Let $\sigma\ll 1$. Then in a successful duplication attempt, the probability
of retaining more than one link is very small (of the order of $\sigma$).
Ignoring it, we conclude that in each successful duplication event, one node
and only one link are added, so when $\sigma\ll 1$ the emerging networks are
trees.

If the degree of the target node is $k$, the probability of the successful
duplication is $1-(1-\sigma)^k$ which approaches $\sigma k$ when $\sigma\ll
1$.  Hence any of the $k$ neighbors of the target node will be linked to the
potentially duplicated node with the same probability $\sigma$.
 
A given node {\bf n} links to the new, duplicated, node in a process which
starts with choosing a neighbor of {\bf n} as the target node. The
probability of that is proportional to the degree $d_n$ of the node {\bf n}.
Then the probability of linking to the node {\bf n} is $\sigma$ (as we already
established) so the probability that the new node links to {\bf n} is
proportional to its degree $d_n$. Thus we recover the standard preferential
attachment model \cite{pref}.  This model exhibits the well-known behavior:
The total number of links is $L=N-1$, and the degree distribution is a
self-averaging quantity peaked around the average,
\begin{equation}
\label{k3}
N_k=\frac{4N}{k(k+1)(k+2)}.
\end{equation}

\section{General case ($0<\sigma<1$)}

We now move on to the discussion of the general case which is only partially
understood.

\subsection{Self-averaging}

Self-averaging of any quantity can be probed by analyzing a relative
magnitude of fluctuations of that quantity. As a quantitative measure we
shall use the ratio of the standard deviation to the average. For the total
number of links,
\begin{equation}
\label{chi-def}
\chi=\frac{\sqrt{\langle L^2\rangle - \langle L\rangle^2}}{\langle L\rangle},
\end{equation}
should vanish in the thermodynamic limit if the total number of links
is the self-averaging quantity. A lack of self-averaging would be extremely
important --- it would imply that a slight deviation in the earlier development
could lead to a very different outcome. Even if $\chi$ vanishes in the
thermodynamic limit, fluctuations may still play noticeable role if $\chi$
approaches zero too slowly.

\begin{figure}[ht] 
 \vspace*{0.cm}
 \includegraphics*[width=0.45\textwidth]{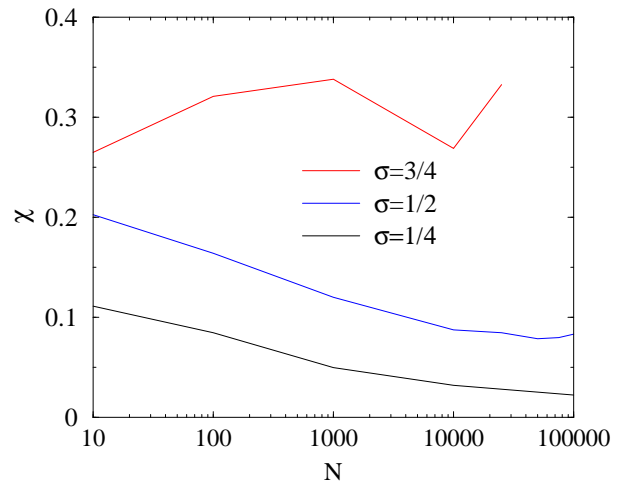}
 \caption{$\chi$ vs. $N$ for (top to bottom)  $\sigma=3/4, 1/2, 1/4$. 
   The total number of nodes is obviously a self-averaging quantity for
   $\sigma=1/4$, apparently also self-averaging for $\sigma=1/2$, and
   evidently non self-averaging for $\sigma=3/4$.} 
\label{chi}
\end{figure}

Simulations (Fig.~\ref{chi}) show that the system is apparently
self-averaging when $\sigma\leq 1/2$. It is somewhat difficult to establish
what is happening in the borderline case $\sigma=1/2$, though we are inclined
to believe that self-averaging still holds.  The self-averaging is evidently
lost at $\sigma=3/4$, and the system is certainly non-self-averaging for
$\sigma=1$ (in this situation $\chi=1/\sqrt{5}$, see
Eqs.~(\ref{Lav})--(\ref{Lav2})).  These findings suggest that in the range
$1/2<\sigma\leq 1$ the total number of links is {\em not} a self-averaging
quantity.

\subsection{Total number of links}

According to the definition of the model, a target node is chosen randomly.
Therefore, the probability that a duplication event is successful, or
equivalently, the average increment of the number of nodes per attempt is 
\begin{equation}
\label{nu}
\Delta N \equiv \nu = \sum_{k\geq 1} n_k\,\left[1-(1-\sigma)^k\right],
\end{equation}
where $n_k=N_k/N$ is a probability for a node to have a degree $k$.
Similarly the increment of the number of links per step is
\begin{equation*}
\Delta L=\sum_{k\geq 1} n_k\,k\sigma
\end{equation*} 
and therefore 
\begin{equation}
\label{LN}
\frac{dL}{dN}=\frac{\sum_{k\geq 1} n_k\,k\sigma}
{\sum_{k\geq 1} n_k\,\left[1-(1-\sigma)^k\right]}.
\end{equation}
The inequality $k\sigma>1-(1-\sigma)^k$ is valid for all $k>1$ and therefore
$dL/dN\geq 1$ implying 
\begin{equation}
\label{L>N}
L\geq N-1.
\end{equation}
This is obvious geometrically as (\ref{L>N}) should hold for any connected
network. 

Using Eq.~(\ref{LN}) we can verify the self-consistency of our
conclusion  (\ref{k3}) derived in the case of $\s=+0$. Substituting
 (\ref{k3}) in (\ref{LN}) we obtain
\begin{equation}
\label{LN1}
\frac{dL}{dN}=1+\s(-\ln \s -1) + {\cal O}[(\s\ln\s)^2].
\end{equation}
It confirms our assumption that for vanishing $\s$, each successful
duplication event increments the number of links by one. 

\begin{figure}[ht] 
  \vspace*{0.cm} \includegraphics*[width=0.45\textwidth]{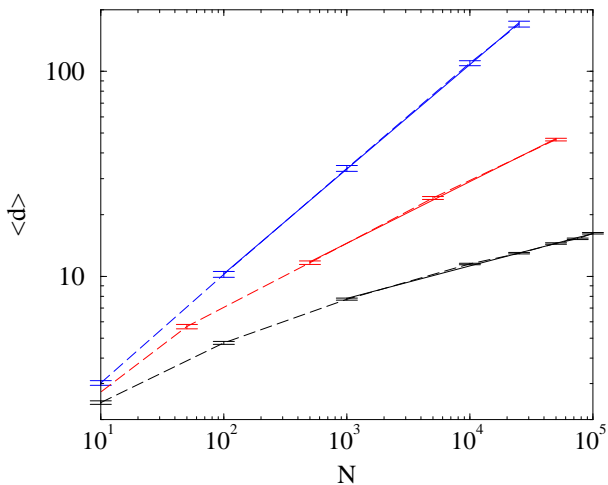}
 \caption{The average node degree $\langle d\rangle$ 
   vs $N$ for (bottom to top, dashed lines ) $\sigma=1/2,5/8,3/4$.  Solid
   lines are corresponding power-law $\langle d\rangle\sim N^{\alpha}$ best
   fits for the large $N$ parts of the plots: $\alpha(\s=1/2)\approx 0.16$,
   $\alpha(\s=5/8)\approx 0.30$, $\alpha(\s=3/4)\approx 0.51$. The results
   are averaged over 100 network realizations. }
\label{large}
\end{figure}

To analyze the growth of $L$ versus $N$, we use the definition (\ref{nu}) of
$\nu$, an identity $2L=\sum kN_k$, and re-write (\ref{LN}) as
\begin{equation}
\label{LN-eq}
\frac{dL}{dN}=\frac{2\sigma}{\nu}\,\frac{L}{N},
\end{equation}
which leads to an algebraic growth $L\sim N^{2\sigma/\nu}$.  Noting that
$\nu$ cannot exceed one (this follows from (\ref{nu}) and the sum rule $\sum
n_k=1$) we conclude that growth is certainly super-linear when $\sigma>1/2$.
Hence the average degree $\langle d\rangle=\sum_{k\ge 1} kn_k=2L/N$
diverges with system size algebraically, $\langle d\rangle \sim N^{\alpha}$
with $\alpha=2\sigma/\nu-1>0$. Since the average degree grows indefinitely,
the probability of the failure to inherit at least one link approaches zero, 
that is $\nu\to 1$ as
$N\to \infty$. Therefore we anticipate that asymptotically $L\sim N^{2\s}$
and $\langle d\rangle \sim N^{\alpha}$ with $\alpha=2\sigma-1>0$. These
expectations agree with simulations fairly well (Fig.~\ref{large}).  For
instance when $\s=3/4$, the predicted exponent $\alpha=1/2$ is close to the
fitted one, $\alpha=0.51$ (Fig.~\ref{large}). The agreement is worse when $\s$
approaches $\s=1/2$; the predicted exponent  for $\s=5/8$
$\alpha=1/4$ is notably smaller than $\alpha_{\rm numer}\approx 0.3$. 

\begin{figure}[ht] 
  \vspace*{0.cm} \includegraphics*[width=0.45\textwidth]{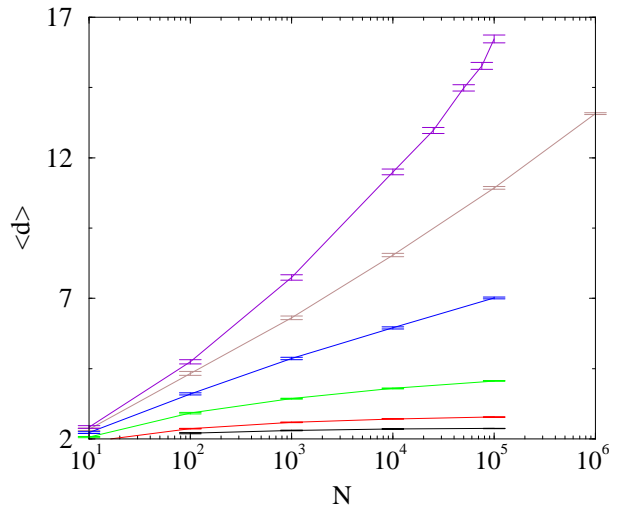}
 \caption{The average node degree $\langle d\rangle$ 
   vs $N$ in the self-averaging regime.  $\sigma=1/16,1/8,1/4,3/8,0.45,1/2$
   (bottom to top). The results are averaged over 100 network realizations. }
\label{small}
\end{figure}

In the range $\sigma\leq 1/2$, we cannot establish on the basis of
Eq.~(\ref{LN-eq}) alone whether the growth is super-linear or linear (the
growth is at least linear as it follows from the lower bound (\ref{L>N})).
The average node degree $\langle d\rangle$ grows with $N$ but apparently
saturates when $\sigma$ is close to zero (see Fig.~\ref{small}). For
$\s\approx 0.3 - 0.4$ the average degree seems to grow logarithmically, that
is $L(N) \sim N\ln N$.  For $\s=1/2$ the growth of $\langle d\rangle$ is
super-logarithmical (see Fig.~\ref{small}) and can be fitted both by $\langle
d\rangle \sim (\ln N)^{\beta}$ with $\beta \approx 2$, or by a power-law
$\langle d\rangle \sim N^{\alpha}$ with a fairly small exponent
$\alpha(1/2)\approx 0.16$.

Hence, taking into account the simulation results and limiting cases
considered earlier, the behavior of $L$ can be summarized as follows:
\begin{equation}
\label{LNb}
L\sim 
\begin{cases}
N^{2\sigma}                    & {\rm for} \quad 1/2<\sigma\leq 1;\cr
N \ln N                   & {\rm for} \quad \s^*\leq \sigma < 1/2;\cr
N                              & {\rm for} \quad 0<\sigma<\s^*;\cr
\end{cases}
\end{equation}
Numerically it appears that $\s^* \approx 0.3 - 0.4$. In the next subsection
we will demonstrate that $\s^*=e^{-1}=0.367879\ldots$.

\subsection{Degree distribution}

A rate equation for the degree distribution is derived in the same manner as
Eq.~(\ref{LN}):
\begin{equation}
\label{NkN}
\nu\,\frac{dN_k}{dN} =\sigma\left[(k-1)n_{k-1}-kn_k\right]+m_k
\end{equation}
Here we have used the shorthand notation 
\begin{equation}
\label{mk}
m_k=\sum_{s\geq k} n_s
{s\choose k}\sigma^k (1-\sigma)^{s-k}
\end{equation}
for the probability that the new node acquires a degree $k$. The general term
in the sum on the right-hand side of Eq.~(\ref{mk}) describes duplication
event in which $k$ links remains and $s-k$ links are lost due to divergence. 

Summing both sides of (\ref{NkN}) over all $k\geq 1$ we obtain $\nu$ on the
left-hand side. On the right-hand side, only the second term contributes to
the sum and also gives the same $\nu$:
\begin{eqnarray*}
\sum_{k\geq 1} m_k&=&\sum_{s\geq 1} n_s
\sum_{k=1}^s {s\choose k}\sigma^k (1-\sigma)^{s-k}\\
&=&\sum_{s\geq 1} n_s[1-(1-\sigma)^s]=\nu, 
\end{eqnarray*}
where the second line was derived using the binomial identity.
Similarly, multiplying (\ref{NkN}) by $k$ and summing over all $k\geq 1$ we
recover (\ref{LN-eq}). These two checks show consistency of (\ref{NkN}) with
the growth equations, introduced earlier.

\begin{figure}[ht] 
 \vspace*{0.cm}
 \includegraphics*[width=0.45\textwidth]{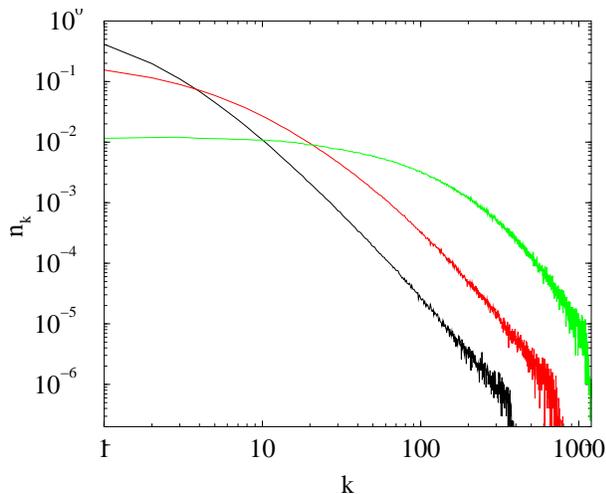}
 \caption{The degree distribution $n_k$ vs. $k$ for (bottom to to top)
  $\sigma=1/4$, $\sigma=1/2$, and $\sigma=3/4$. The size of the network is
  $N=10^5$ 
for $\sigma=1/4$, $N=5\times 10^4$ for $\sigma=1/2$, and 
$N=10^4$ for $\sigma=3/4$. The results are averaged over 100 realizations.}
\label{degree}
\end{figure}

Since $\nu$ depends on all $n_k$, see (\ref{nu}), Eqs.~(\ref{NkN}) are
non-linear.  However, the observations made in the previous subsection allow
us to approximate, for any given $\s$, $\nu$ as parameter, thus ignoring its
possible very slow dependence on $N$.  Resulting linear Eqs.~(\ref{NkN}) are
still very complicated: If we assume that $k\gg 1$ and employ the continuous
approach, we still are left with a system of partial differential equations
with a non-local ``source'' term $m_k$. Fortunately, the summand in $m_k$,
that is $g(s,k)={s\choose k}\sigma^k (1-\sigma)^{s-k}$, is sharply peaked
around $s\approx k/\sigma$ \cite{korea}.  Hence we can replace $\sum_{s\geq
  k}n_sg(s,k)$ by $n_{k/\sigma}\sum_{s\geq k}g(s,k)\equiv
\sigma^{-1}n_{k/\sigma}$ \cite{identity}, and Eqs.~(\ref{NkN}) become
\begin{equation}
\label{NkN-pde}
\nu\,N\,\frac{\partial}{\partial N}\, N_k +\sigma\,
\frac{\partial}{\partial k}\,k N_k=\sigma^{-1}N_{k/\sigma}
\end{equation}

Still, the analysis of (\ref{NkN-pde}) is hardly possible without knowing the
correct scaling. Figure \ref{degree} indicates that the form of the degree
distribution varies with $\s$ significantly. We will proceed (separately for
$0<\sigma<1/2$ and $1/2<\sigma<1$) by guessing the scaling and trying to
justify the consistency of the guess.

\subsubsection{$0<\sigma<1/2$}

{\em Assuming} the simplest linear scaling $N_k\sim N$ we reduce
Eq.~(\ref{NkN-pde}) to
\begin{equation}
\label{nk}
2n_k+\frac{d}{dk}\,kn_k=\sigma^{-2}\,n_{k/\sigma}.
\end{equation}
We also used $\nu=2\sigma$, which is required to assure that $L\sim N$
\cite{comment} is consistent with (\ref{LN-eq}).  Plugging $n_k\sim
k^{-\gamma}$ into (\ref{nk}) we obtain
\begin{equation}
\label{gamma}
\g=3-\s^{\g-2}.
\end{equation}
\begin{figure}[ht] 
 \vspace*{0.cm}
 \includegraphics*[width=0.45\textwidth]{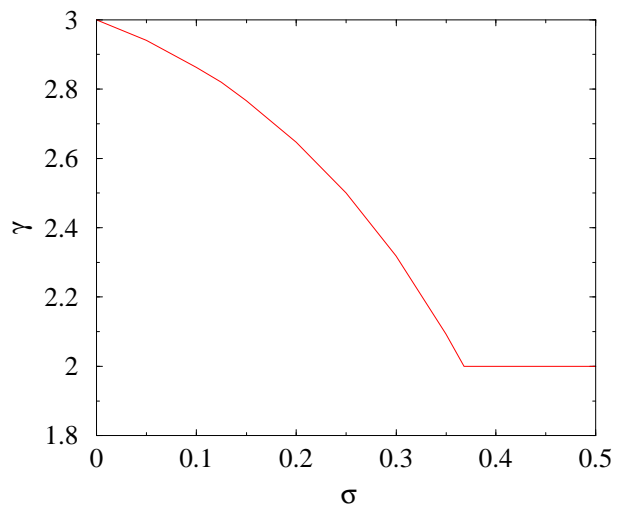}
 \caption{The degree distribution exponent $\g(\s)$ from Eq.~(\ref{gamma}).}
\label{fig_g}
\end{figure}
This equation has two solutions: $\gamma=2$ and a non-trivial solution
$\gamma(\sigma)$ which depends on $\sigma$. The second solution
$\gamma(\sigma)$ decreases from $\gamma(0)=3$ to $\gamma(1/2)=1$.  The two
solutions coincide at $\sigma^*=e^{-1}=0.367879$.  The sum $\sum kn_k$
converges when $\gamma>2$, and the total number of links grows linearly,
$L\sim N$. Apparently the appropriate solution is the one which is larger:
For $\sigma<e^{-1}$ the exponent is $\gamma(\sigma)$, while for
$\sigma>e^{-1}$ the exponent is $\gamma=2$, Fig.~\ref{fig_g}. In the latter
case,
\begin{equation*}
\sum_{k<k_{\rm max}} kn_k\sim \sum_{k<k_{\rm max}} k^{-1}\sim \ln k_{\rm
  max}\sim \ln N
\end{equation*}
and therefore the total number of links grows as $N\ln N$. 

\begin{figure}[ht] 
 \vspace*{0.cm}
 \includegraphics*[width=0.45\textwidth]{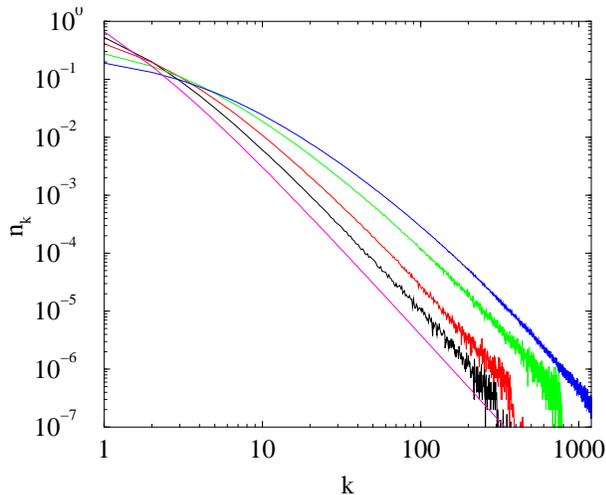}
 \caption{$n_k$ vs $k$ for the network of size
   $N=10^5$ in the self-averaging regime. 
   $\sigma= +0, 1/8, 1/4, 3/8, 0.45$ (bottom to top). The result for
   $\s=+0$ is the exact solution (\ref{k3}), simulation data is averaged over
   100 realizations.  The corresponding analytical
   predictions for the exponent are $\gamma(\s=1/8)=2.817187$, $\gamma(\s=1/4)
   =5/2$, and $\gamma(\s=3/8)=\gamma(\s=0.45)=2$.}
\label{less}
\end{figure}

Simulations show that for small $\s$ the degree distribution $n_k$ has indeed
a fat tail (see Fig.~\ref{less}). The agreement with the theoretical
prediction of the algebraic tail is very good when $\sigma=1/8$
(Eq.~(\ref{gamma}) gives $\gamma=2.817187$ while numerically $\gamma_{\rm
  numer}\approx 2.82$), not so good when $\sigma=1/4$ ($\gamma=5/2$ vs.
$\gamma_{\rm numer}\approx 2.7$), and fair at best for $\sigma=3/8$.

Thus we explained the growth law (\ref{LNb}). We also arrived at the
theoretical prediction of $\s^*$ which reasonably well agree with simulation
results.  Due to the presence of logarithms, the convergence is extremely
slow and better agreement will be probably very hard to achieve. Finally we
note that the behaviors $L\sim N\ln N$ and $n_k\sim k^{-2}$ arise in a
surprisingly large number of technological and social networks (see
\cite{KR}  and references therein). 

\subsubsection{$1/2<\sigma<1$}

\begin{figure}[ht] 
 \vspace*{0.cm}
 \includegraphics*[width=0.45\textwidth]{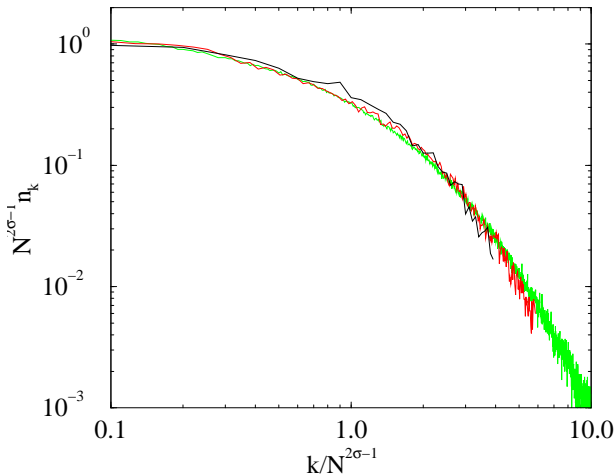}
 \caption{Scaling 
   of the degree distribution in the networks of
   $N=100$, $N=1000$, and $N=10000$ nodes with $\sigma=3/4$.}
\label{more}
\end{figure}

The growth law (\ref{LNb}) suggests an introduction of a scaling form
$N_k=N^{2-2\sigma}F(x)$ with $x=k/N^{2\sigma-1}$. Then the sum rules $\sum
N_k=N$ and $\sum kN_k\sim N^{2\sigma}$ are manifestly satisfied (provided that
the scaling function $F(x)$ falls off reasonably fast for $x\to\infty$).
Simulation results (see Fig.~\ref{more}) are in a good agreement with above
scaling form.

\section{Conclusions}

We have shown that a simple one-parameter duplication-divergence network
growth model well approximates realistic protein-protein networks.
Table~\ref{tab_uno} summarizes how the major network features
(self-averaging, evolution of the number of links $L(N)$, the degree
distribution $n_k$) change when the link retention probability $\s$ varies.
\begin{table*}
\begin{ruledtabular}
\begin{tabular}{cccc}
$\s$ & self-averaging & $L(N)$ & $n_k$ \\
\hline
\hline
$\s=1$ &  No & $N(N+1)/6$ & $2(N-k)/[N(N-1)]$\\
\hline
$1/2<\s<1$ &  No & $\sim N^{2\s-1}$ & $\sim N^{1-2\sigma}F(k/N^{2\sigma-1})$\\
\hline
$e^{-1}\leq\s < 1/2$ & Yes & $\sim N \ln N $ & probably $\sim
k^{-2}$\\ 
\hline
$0<\s<e^{-1}$ & Yes & $\sim N$ & $\sim k^{-\g(\s)}$\\
\hline
$\s=+0$ & Yes & $N-1$ & $4/[k(k+1)(k+2)]$\\
\end{tabular}
\end{ruledtabular}
\caption{\label{tab_uno}
The behavior of the duplication-divergence network for different values of
probability to inherit a link $\s$. Here $L(N)$ is the average number of links for given
number of nodes $N$, $n_k$ the average fraction of nodes of degree $k$, and the exponent 
$\g(\s)>2$ is defined by equation $\g=3-\s^{\g-2}$.}
\end{table*}

Two most striking features of duplication-divergence networks are the lack of
self-averaging for $\s>1/2$ and extremely slow growth of the average degree
for $\s<1/2$. These features have very important biological implications: The
lack of self-averaging naturally leads to a diversity between the grown
networks and the slow degree growth preserves the sparse structure of the
network. Both of these effects occur in wide ranges of parameter $\s$ and
therefore are robust --- it is hard to expect that nature would have been
able to fine-tune the value of $\s$ if it were not so.

Our findings indicate that in the observed protein-protein networks $\s
\approx 0.4$, so biologically-relevant networks seem to be in the
self-averaging regime.  One must, however, take the experimental
protein-protein data with a great degree of caution: It is generally
acknowledged that our understanding of protein-protein networks is quite
incomplete. Usually, as the new experimental data becomes available, the
number of links and the average degree in these network increases. Hence the
currently observed degree distributions may reflect not any intrinsic
property of protein-protein networks, but a measure of an incompleteness of
our knowledge about them.  Therefore a possibility that the real
protein-protein networks are not (or have not been at some stage of the
evolution) self-averaging is not excluded.

\begin{figure}
\includegraphics[width=.45\textwidth]{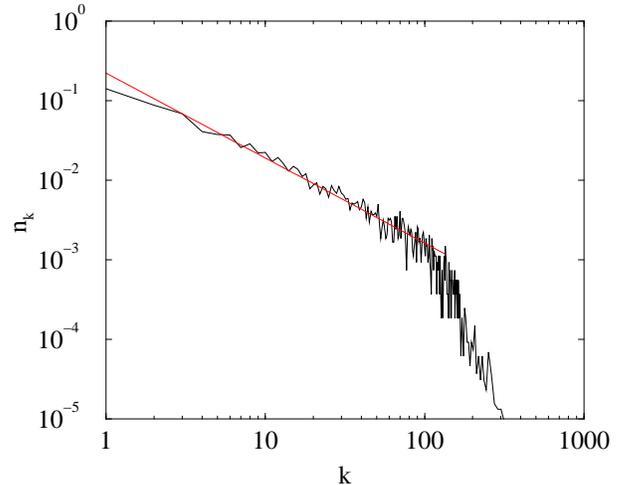}
\caption{\label{fig_vM} Using a multitude of direct and indirect 
methods, von Mering et al
\cite{vM} predicted 78928 links between 5397 yeast proteins which produces a
network with 
the  average degree $\langle d \rangle \approx 29.2$. A power-law fit to this
degree distribution has the exponent
$\g\approx1.1$. 
}  
\end{figure}

It has been suggested that randomly introduced links (mutations) must
compliment the inherited ones to ensure the self-averaging and existence of
smooth degree distribution \cite{pastor}.  While a lack of random linking
does affect the fine structure of the resulting network, we have observed
that the major features like self-averaging, growth law, and degree
distribution are rather insensitive to whether random links are introduced or
not, provided that the number of such links is significantly less than the
number of inherited ones.  We performed a number of simulation runs where
links between a target node and its image were added at each duplication step
with a probability $P_d$.  Introduction of such links is the most direct way
to prevent partitioning of the network into a bipartite graph (see
\cite{korea}). In other words, without such links the target and duplicated
nodes are never directly connected to each other.  We observed that for
reasonable values of $P_d<0.1$ (in the observed yeast, fly, and human
protein-protein networks $P_d$ never exceeds this value) the results remain
unaffected.  Apparently, without randomly introduced links, the network
characteristics establish themselves independently in every subset of
vertices duplicated from each originally existing node.  We leave more
systematic study of the effects of mutations as well as of the more symmetric
divergence scenarios (when links may be lost both on the target and
duplicated node) for the future.

Many unanswered questions remain even in the realm of the present model.  For
instance, little is known about the behavior of the system in the borderline
cases of $\sigma=1/2$ and $\sigma=e^{-1}$. One also wants to understand
better the tail of the degree distribution in the region $\sigma\geq e^{-1}$
where $L(N)$ follows unusual scaling laws. It will be also interesting to
study possible implications of these results for the probabilistic urn models
\cite{JK}.

\section{acknowledgment}
The authors are thankful to S.~Maslov, S.~Redner, and M.~Karttunen
for stimulating discussions. 
This work was supported by 1 R01 GM068954-01 grant from NIGMS.


\begin{thebibliography}{99}

\bibitem{japan}  S.~Ohno, {\em Evolution by gene duplication} 
                 (Springer-Verlag, New York, 1970).

\bibitem{bio}    J.~S.~Taylor and J.~Raes, 
                 Annu.\ Rev.\ Genet. {\bf 9}, 615--643 (2004). 

\bibitem{wag}    A. Wagner,   
                 Proc. R. Soc. Lond. B {\bf 270}, 457--466 (2003).

\bibitem{wagas}  G.~C.~Conant and A.~Wagner,
                 Genome Research {\bf 13}, 2052 (2003). 

\bibitem{korea}  J.~Kim, P.~L.~Krapivsky, B.~Kahng, and S.~Redner,
                 Phys.\ Rev.\ E. {\bf 66}, 055101 (2002).

\bibitem{chung}  F.~Chung, L.~Lu, T.~G.~Dewey, and D.~J.~Galas,
                 J. Comput.\ Biol. {\bf 10}, 677--687 (2003).

\bibitem{pastor} R.~V.~Sol\'e, R.~Pastor-Satorras, E.~D.~Smith, and
                 T.~Kepler, Adv.\ Complex\ Syst. {\bf 5}, 43 (2002).

\bibitem{bb}     M.~Bauer and D.~Bernard, 
                 J. Stat.\ Phys. {\bf 111}, 703--737 (2003).

\bibitem{cb}     C.~Coulomb and M.~Bauer, 
                 Eur.\ Phys.\ J. B. {\bf 35}, 377--389 (2003).

\bibitem{kd}     P.~L.~Krapivsky and B.~Derrida,  
                 Physica A {\bf 340}, 714--724 (2004).  

\bibitem{ag}     http://www.ariadnegenomics.com/.

\bibitem{medscan1}Svetlana Novichkova, Sergei Egorov and Nikolai
                  Daraselia, Bioinformatics {\bf 19}, 1699 (2003).

\bibitem {pwa}   http://www.ariadnegenomics.com/products/pathway.html.

\bibitem{raval}  A.~Raval, 
                 Phys.\ Rev.\ E. {\bf 68}, 066119 (2003).

\bibitem{pol}    F.~Eggenberger and G.~P\'olya, 
                 Zeit.\ Angew.\ Math.\ Mech. {\bf 3}, 279 (1923). 

\bibitem{JK}     N.~L.~Johnson and S.~Kotz, 
                 {\em Urn Models and their Applications} (Wiley, New York, 1977). 
 
\bibitem{life}   K.~Sigmund, {\em Games of Life}
                 (Oxford University Press, Oxford, 1993). 

\bibitem{BP}     A.~Bagchi and A.~K.~Pal, 
                 SIAM J. Alg.\ Disc.\ Meth. {\bf 6}, 394 (1985). 

\bibitem{A}      D.~Aldous, B.~Flannery, and J.~L.~Palacios,
                 Prob.\ Eng.\ Infor.\ Sci. {\bf 2}, 293 (1988). 

\bibitem{KMR}    S.~Kotz, H.~Mahmoud, and P.~Robert, 
                 Stat.\ Probab.\ Lett. {\bf 49}, 163 (2000).  

\bibitem{Fla}    P.~Flajolet, J.~Cabarro, and H.~Pekari, Ann.\ Prob. (2004). 

\bibitem {pref}  H.~A.~Simon, Biometrica {\bf 42}, 425 (1955); Infor. Control 
                 {\bf 3}, 80 (1960).  

\bibitem{identity} 
  We used the identity $\sum_{s\geq k}{s\choose k}\sigma^k
  (1-\sigma)^{s-k}=\sigma^{-1}$.

\bibitem{comment}The growth law $L\sim N\ln N$ agrees with (\ref{LN-eq}) if
                 the convergence of $\nu$ to $2\s$ is logarithmically slow: 
                 $2\s-\nu(N)\sim (\ln N)^{-1}$.

\bibitem{KR}     P.~L.~Krapivsky and S.~Redner,
                 cond-mat/0410379. 

\bibitem {vM}    C. von Mering {\em et al.}, Nature {\bf 417}, 399 (2002).

\end{thebibliography}
\end{document}